# Latency Measurement of 100 km Fiber Using Correlation-OTDR

Florian Azendorf
Advanced Technology
ADVA Optical Networking SE
Meiningen, Germany
fazendorf@advaoptical.com

Annika Dochhan
Advanced Technology
ADVA Optical Networking SE
Meiningen, Germany
adochhan@advaoptical.com

Bernhard Schmauss
Optical Radio Technology and Photonics
University Erlangen
Erlangen, Germany
bernhard.schmauss@fau.de

Michael Eiselt
Advanced Technology
ADVA Optical Networking SE
Meiningen, Germany
meiselt@advaoptical.com

*Abstract*—**By means of C-OTDR (Correlation - Optical Time Domain Reflectometry), we measured the latency of 100 km fiber with an accuracy of a few picoseconds. Based on iterating 49 measurements, we calculated a standard deviation of 12 ps between the round-trip latency values. To verify the reflection measurements, we used a single pass setup without reflector, which showed a maximum difference of only 11 ps.**

## I. Introduction

For next generation networks, fiber latency becomes a critical parameter, especially for synchronization applications over optical networks. New protocols, which are based on IEEE 1588 PTP, accept an asymmetry between the propagation directions between master and slave of ~2–5 ns per span. If the asymmetry is higher, synchronization errors can occur. Therefore, it is necessary to know the behavior of the fiber latency due to environmental impacts, like temperature variations, which lead to a change of the refractive index of the fiber. The TDC (Temperature Delay Coefficient) describes the latency impact due to temperature variations of a single-mode fiber. A typical value is $7 \cdot 10^{-6}$/K [1]. Therefore, it is necessary to monitor and to compensate the differential fiber latency of the link. We investigated a correlation OTDR measurement method for fiber latency, which provides an accuracy of a few picoseconds. To cover a wide range of network applications, we targeted a fiber length of 100 km. As compared to previous publications, where we measured shorter lengths of up to 8.5 km [1–3], we adapted the measurement setup with respect to transmission rate, pattern sequence and length, bandwidth, and sampling rate of the oscilloscope at the receiver. Instead of 10 Gbit/s transmission rate we now used 2 Gbit/s to reduce the impact of chromatic dispersion after propagation of the reflected signal over up to 200 km. The bandwidth of the receiver was set to 2 GHz, and a sample rate of 10 GS/s was used. To improve the SNR, we used Golay sequences instead of PRBS [4]. By using two complementary Golay sequences and adding the correlation functions, correlation sidelobes can be reduced, while at the same time the correlation amplitude is doubled.

## II. Measurement setup

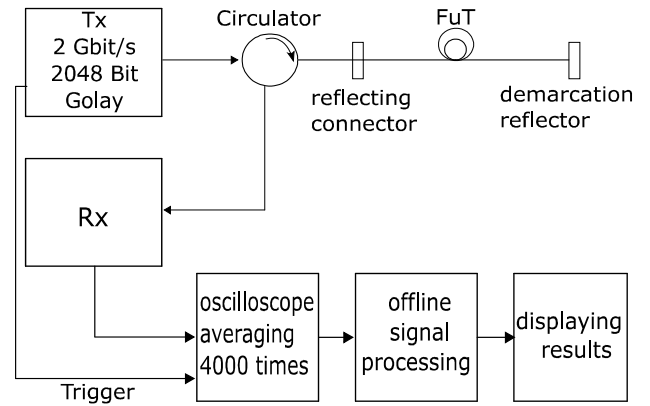

Fig. 1. Measurement setup FuT fiber under test

The measurement setup is shown in Fig. 1. A continuous wave laser with a wavelength of 1550 nm was used. The light was modulated, using a Mach-Zehnder modulator, with a 2-Gbit/s, 2048-Bit Golay sequence, followed by zeros, to get a packet with a length of 1 ms to cover the round-trip time of a 100-km fiber link. The signal was sent into the fiber under test (FuT) via a circulator. To mark the beginning of the FuT, we used a connector with an air gap as reference reflector. Two 50-km single mode fiber spools were used as FuT. A reflector with a reflection factor of R = 95% for the probe wavelength was added at the end of the second spool. The reflected light was received after the circulator with a PIN/TIA receiver and recorded on a real time oscilloscope with a sampling rate of 10 GS/s. The oscilloscope was synchronized by the burst trigger of the Golay source. During the observation time, the oscilloscope recorded and averaged 4000 traces. The first 3 µs of the received and averaged signal, showing the reflection of the Golay sequence by the reference reflection is shown in Fig 2. Similarly, the averaged reflection coming from the fiber end after a round trip of 200 km is shown in Fig. 3. The Golay sequence cannot be directly seen in the noise, anymore.

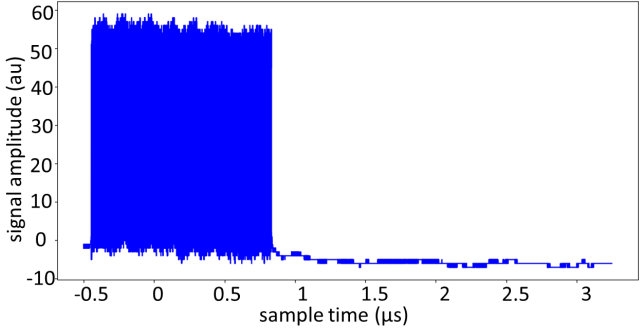

Fig. 2. Averaged signal showing the Golay sequence reflected by the reference reflection

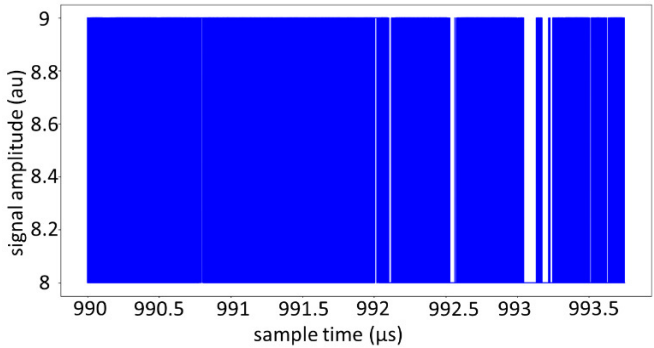

Fig. 3. Averaged signal reflected from fiber end

After averaging, the received signals were correlated with the respective transmitted Golay sequences. Fig 4. shows the correlation of the signal in Fig. 3 with the Golay sequences. Here, the sum of the correlation results with the two complementary sequences are shown. It can be seen that the correlation yields a good reflection peak, even if the SNR is low as shown in Fig. 3.

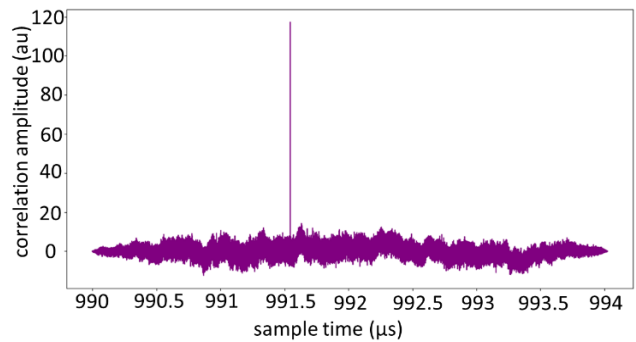

Fig. 4. Correlated sum of Golay sequences showing reflection peak at the fiber end

To find the reflection time with improved accuracy, seven points around the highest reflection peak were fitted to a raised cosine function, as shown in Fig. 5. While the sampling rate would yield a resolution of 100 ps, the fitting improved the accuracy to a few picoseconds.

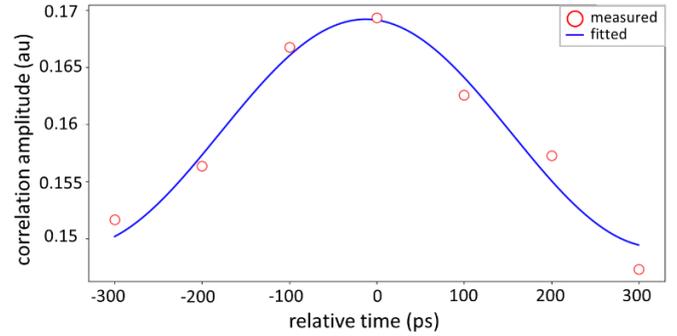

Fig. 5. Fitting of the reflection peak

### III. RESULTS

#### A. Repeatability over time

With the described setup, the round-trip latency of the 100-km fiber was measured 49 times over a period of three hours. Fig. 4 shows the evolution of the measured values. The round-trip latency increased over time by approximately 1.1 ns, corresponding to a relative change of 1.1 ppm. This change can be explained by a temperature change of 0.16 K, which is typical for the laboratory environment during a period between 9 am and noon. To exclude the latency changes from the repeatability estimation, a 4$^{th}$ order polynomial function over time was fit to the measurement results and the difference between the measurement results and the fit was calculated. The resulting standard deviation was 12 ps, pointing to a repeatability of the measurement of this magnitude.

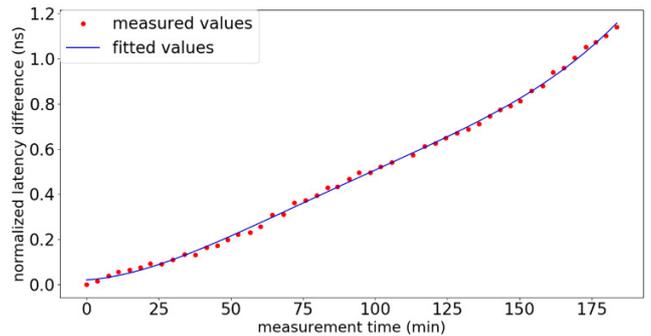

Fig. 6. Latency measurements over time (offset by 990.537 µs) with 4$^{th}$ order polynomial fit

## B. Comparison with single-pass measurement

To verify the reflection measurement results, we used a single pass method in a setup similar to Fig. 1. The signal was fed into the fiber from one end and detected at the other end. A reference length was measured after disconnecting the FuT at the two reflective connectors. The difference between the round-trip time latency, measured in the reflective setup and divided by two, and the single pass setup was 11 ps and 1 ps, respectively, in two consecutive pairs of measurements. This indicates a high accuracy of the measurement despite the different chromatic dispersion values experienced by the single-pass and the reflected signals.

A second factor impacting the accuracy of the measurement is the accuracy of the sampling frequency, provided by the oscilloscope. A clock frequency error would affect both, the single-pass and the round-trip measurements. Based on the data sheet accuracy of the used oscilloscope, we estimate the clock error to be smaller than ±0.5 ppm, resulting in an absolute latency error of less than 250 ps. This is much more than the error due to the measurement method. However, when only the differential latency between two fibers is of interest, as for most synchronization applications, the same sampling frequency is used for both fibers and the impact of the clock error is much smaller.

## IV. CONCLUSION

We have demonstrated that, using the C-OTDR method, it is possible to measure the latency of a 100-km fiber link with a high accuracy and repeatability, which was better than 12 ps. Comparing with a single-pass measurement, we can calculate a difference of less than 11 ps. A major error impact is due the inaccuracy of the measurement sample frequency, which can be overcome by using a high-precision clock. It is therefore possible to use this measurement method to accurately monitor the fiber latency.

## V. ACKNOWLEDGEMENT

This project has received funding from the European Union´s Horizon 2020 research and innovation programme under grant agreement No 762055 (blueSpace Project)